\documentclass[amssymb,amsmath,showpacs,aps,pra,floatfix]{revtex4-1}
\usepackage{color,graphicx,bm}
\usepackage{overpic}
\usepackage{upgreek}
\usepackage{graphicx}

\def\beq{\begin{equation}}
\def\eeq{\end{equation}}

\def\eps{\epsilon}

\begin{document}

\title{Impact of grain boundary characteristics on thermal transport in polycrystalline graphene: Analytical results}
\author{S.E. Krasavin and V.A. Osipov}

\vspace{0.3 cm}

\affiliation{
Joint Institute for Nuclear Research,\\
Bogoliubov Laboratory of Theoretical Physics\\
141980 Dubna, Moscow region, Russia\\
e-mail: krasavin@theor.jinr.ru osipov@theor.jinr.ru, 
}
\pacs {PACS numbers: 65.80.Ck, 61.72.Mm, 63.20.kp}


\begin{abstract}

The effect of grain boundary (GB) structure, size and shape on thermal conductivity of polycrystalline graphene is studied in the framework of the deformation potential approach. Precise analytical expressions  for the phonon mean free path (MFP) are obtained within the Born approximation. We found exactly two types of behavior in the long-wavelength limit: MFP varies as $\omega ^{-1}$ for open GBs of any shape while it behaves as $\omega ^{-3}$ for closed configurations (loops). In the short-wavelength limit MFP tends to a constant value for any configuration. Oscillatory behavior is observed for all GBs which indicates that they serve as diffraction grating for phonons. This property is also inherent in GBs with irregularities caused by partial disclination dipoles. The thermal conductivity is calculated in the framework of Callaway's  approach with all main sources of phonon scattering taken into account. Reduction of the heat conductivity with decreasing grain size is obtained in a wide temperature range. Most interesting is that we found a marked decrease in the thermal conductivity of polycrystalline graphene containing  GBs with changes in their misorientation angles.

\end{abstract}
\maketitle
 


\section{Introduction}

The large-scale graphene films synthesized by CVD method are typically polycrystalline. This must be taken into account in any practical applications of graphene, such as sensors, detectors, etc. Indeed, it was found that grain boundaries (GBs) have a significant effect on electrical~\cite{yuz}, mechanical~\cite{gran} and other physical properties of graphene. It is of great interest to study the influence of GBs on the thermal conductivity of graphene. 

Two decades ago we suggested a model based on wedge disclination dipoles (WDDs) which allowed us to describe the phonon scattering by grain boundary of finite length~\cite{jpcm98}. Actually, this model relies on the fact that grain boundaries can be described in some cases by rotational rather than translational defects and therefore it is more natural to model them by dislinations~\cite{Li}. We used the fact that the far strain fields caused by WDDs agree with those from finite walls of edge dislocations. This makes it possible to extend the well-known Klemens model to take into account the finiteness of the boundary. In particular, we found that the biaxial WDD with nonskew axes of rotation (BWDD) shows a very specific behavior when the phonon mean free path (MFP) $l$ exhibits a crossover from $l\sim\omega^{-1}$ at low frequencies to a constant value with increasing $\omega$. This leads to the corresponding crossover of the thermal conductivity, $\kappa$,  from $\kappa\sim T^2$ to $\kappa\sim T^3$, which is of importance in the description of occasionally observed low-temperature anomalies~\cite{kras3}.

It is interesting that in some modern 2D polycrystalline materials like graphene and phosphorene grain boundaries were found to be constructed of 5-7 rings or, equivalently, 5-7 disclination dipoles~\cite{yuz,rom}. Moreover, these dipoles are nothing else but BWDDs which serve as a basic structural unit of various types of grain boundaries. Placing 5-7 BWDDs in a continuous line one obtains a worm-like GB. A space between 5-7 dipoles is directly related to the misorientation angles of GBs. Thus, the problem arises of describing phonon scattering by such walls made from BWDDs. Much work in this direction has been done within molecular dynamics (MD) simulations. In particular, it has been found in equilibrium MD simulations that the thermal conductivity of ultra-fine grained graphene is one order of magnitude lower than that for pristine graphene~\cite{mort}. Besides, the effect of sharp temperature jump at the grain boundary was observed within non-equilibrium MD simulations~\cite{bagri}. All MD calculations predict a decrease in thermal conductivity with decreasing size of the grain boundary. A similar behavior was revealed using both Landauer-B\"uttiker~\cite{serov} and Boltzmann transport formalism~\cite{aksamiya} where, in addition, the important role of individual properties of GBs was noted.

In this paper, we present exact analytical results for the GB-induced phonon mean free path within the framework of BWDD's model. Proposed calculation scheme allows us to consider any configuration of BWDDs, both straight GBs oriented along some axis (open GB)~\cite{wu1} and any closed GBs when 5-7 dipoles arranged as loops in pristine graphene~\cite{cock}. In so doing we mean that GB can change its direction locally. We analyze the thermal conductivity of polycrystalline graphene using the Callaway formalism. It was previously applied by the authors to describe Stone-Wales (SW) defects in graphene~\cite{kras1} where SW defect has been  considered as a quadrupole of wedge disclinations or, equivalently, a combination of two 5-7 pairs~\cite{ma}.

\section{Model}

Taking into account the 2D character of the phonon transport in graphene, the GB-induced  MFP  is written as~\cite{ziman}
\begin{eqnarray}
l^{-1}=n_{def}\int_{0}^{2\pi}(1-\cos\theta )R(\theta )d\theta
\end{eqnarray}
with $R(\theta )$ being the effective differential scattering width, $\theta $ the scattering angle, and $n_{def}$ the areal density of defects.  We restrict our consideration here to the most important TA and LA acoustic phonon branches. Within the Born approximation $R(\theta )$ takes the form
\begin{eqnarray}
R(\theta )=\frac{kS^{2}}{2\pi \hbar ^{2}v_{\lambda }^{2}}\overline{|\langle {\bf k}|U(r)|{\bf k^{'}}\rangle|^{2}},
\end{eqnarray}
where $S$ is a projected area, $v_{\lambda }$ is the sound velocities of the phonon branches $\lambda $ ($\lambda $=TA,LA),  $U(r)$ is the effective perturbation energy of a phonon due to strain fields caused by the GB, and the bar denotes an averaging procedure over $\alpha$ which defines an angle between the scattering vector $\bf q=\bf k - \bf k^{'}$ and the axis along the GB orientation. 

Assuming that the effect of strain fields leads only to a change in the velocity of sound $v_{\lambda }$ and taking the energy of phonon with frequency $\omega _{\lambda }(k)$ as $\hbar kv _{\lambda }$, the perturbation energy is written as 
\begin{equation}
U(r)=\hbar kv_{\lambda}\gamma_{\lambda }TrE_{ij}^{d}(r).
\end{equation}
Here $TrE_{ij}^{d}(r)$ is the trace of the strain tensor due to GB,  $\gamma _{\lambda }$ is the Gr\"uneisen constant for a given phonon branch $\lambda $, and $\hbar $ is the Planck constant. The strain field caused by GB of any size and shape can be obtained as a sum of strains of 5-7 BWDDs (see Appendix). 
The explicit expression for $R(\theta )$ is found immediately after calculating the Fourier transform of $U(r)$ given by Eq.(3) and Eq.A(3) 
$$
\langle {\bf k}|U(r)|{\bf k^{'}}\rangle =\frac{1}{S}\int d^{2}r U(x,y)\exp\Bigl( iq_{x}x+iq_{y}y\Bigr)
$$
\begin{eqnarray}
=-\frac{4\pi A}{Sq^2}\sum_{n=1}^{p}\Bigl(\exp(iq_{x}x_{n1})\exp(iq_{y}y_{n1})-\exp(iq_{x}x_{n2})\exp(iq_{y}y_{n2})\Bigr),
\end{eqnarray}
where $A=\hbar k v_{\lambda }\gamma _{\lambda }\nu (1-2\sigma )/2 (1-\sigma )$, $q=|k-k'|=2k\sin (\theta/2)$, $\nu=\Omega /2\pi $, and $\Omega$ is the value of the Frank vector.
Finally,  after averaging $|\langle {\bf k}|U(r)|{\bf k^{'}}\rangle|^{2}$  with respect to $\alpha $, we obtain 
$$
R(\theta )=\frac{\pi A^{2}}{\hbar ^{2}v_{\lambda}^24k^{3}\sin ^{4}(\theta/2) }\sum_{n=1}^{p}\sum_{m=1}^{p}\{J_{0}(|2k\sin (\theta/2)\sqrt{((x_{n1}-x_{m1})^2+(y_{n1}-y_{m1}])^2)}|)-
$$
$$
J_{0}(|2k\sin(\theta/2)\sqrt{((x_{n2}-x_{m1})^2+(y_{n2}-y_{m1}])^2)}|)-J_{0}(|2k\sin (\theta/2)\sqrt{((x_{n1}-x_{m2})^2+(y_{n1}-y_{m2}])^2)}|)
$$
\begin{eqnarray}
+J_{0}(|2k\sin (\theta/2)\sqrt{((x_{n2}-x_{m2})^2+(y_{n2}-y_{m2}])^2)}|)\},
\end{eqnarray}
where $J_{0}(|2k\sin (\theta/2)\sqrt{((x_{ni}-x_{mj})^2+(y_{ni}-y_{mj}])^2)}|)$ is the Bessel function of the first kind.
Accordingly, after integration over $\theta $ in Eq.(1), the phonon MFP takes the form
$$
l^{-1}_{GB,\lambda}=\frac{n_{GB}\nu ^2D^2\pi }{4 k}\sum_{n=1}^{p}\sum_{m=1}^{p}\Biggl(-\mathfrak{S}\biggl(k,x_{n2},x_{m1},y_{n2},y_{m1}\biggr)-\mathfrak{S}\biggl(k,x_{n1},x_{m2},y_{n1},y_{m2}\biggr)
$$
\begin{eqnarray}
+\mathfrak{S}\biggl(k,x_{n1},x_{m1},y_{n1},y_{m1}\biggr)+\mathfrak{S}\biggl(k,x_{n2},x_{m2},y_{n2},y_{m2}\biggr) \Biggr),
\end{eqnarray}
where $D=\gamma _{\lambda }(1-2\sigma )/(1-\sigma )$ and $n_{GB}$ is the areal density of GBs. The explicit expression for $\mathfrak{S}\biggl(k,x_{n1},x_{m2},y_{n1},y_{m2}\biggr)$ is given by Eq.(A4) in Appendix.   

\subsection{Phonon mean free path}

Let us consider the case $p=2$ and put dipoles at points  $(x_{1i},y_{1i})=(\pm L/2+\delta,L/2\pm \delta )$, $(x_{2i},y_{2i})=(\mp L/2-\delta,-L/2\mp \delta )$. Such an arrangement corresponds to the SW defect.
One obtains (see also Ref.~\cite{kras1})
\begin{eqnarray}
l^{-1}_{SW,\lambda}=2\pi ^{2}kn_{sw}D^2\nu ^2\Biggl(2\tilde L^2\biggl(J_{0}^{2}(k\tilde L)+J_{1}^{2}(k\tilde L)-
J_{0}(k\tilde L)J_{1}(k\tilde L)/k\tilde L\biggr)- \\ \nonumber
\sum_{n=1}^2 L_n^2\biggl(J_{0}^{2}(\sqrt{2}kL_n)+J_{1}^{2}(\sqrt{2}kL_n)-
J_{0}(\sqrt{2}kL_n)J_{1}(\sqrt{2}kL_n)/\sqrt{2}kL_n\biggr)\Biggr),
\end{eqnarray}
where  $n_{SW}$ is the areal density of SW defects,  $\tilde L=\sqrt{L^2+4\delta ^2}$, $L_{1}=L+2\delta $, $L_{2}=L-2\delta $. 

For rectilinear GBs, there are two characteristic sizes: a dipole arm $L$ and a distance between dipoles $h$. Having dipoles in the points $(x_{11},y_{11})=(L, 0)$, $(x_{12},y_{12})=(0,0)$, $(x_{21},y_{21})=(2L+h, 0)$, $(x_{22},y_{22})=(L+h, 0)$ we obtain  

\begin{eqnarray}
l^{-1}_{GB,\lambda}=\pi ^{2}kn_{GB}D^2\nu ^{2}\Biggl(\sum_{n=1}^3 Z_n^2\biggl(J_{0}^{2}(kZ_n)+J_{1}^{2}(kZ_n)-
J_{0}(kZ_n)J_{1}(kZ_n)/kZ_n\biggr)-\\ \nonumber
 2Z_{4}^2\biggl(J_{0}^{2}(kZ_{4})+J_{1}^{2}(kZ_{4})-
J_{0}(kZ_{4})J_{1}(kZ_{4})/kZ_{4}\biggr)\Biggr),
\end{eqnarray}
where $Z_{1}=\sqrt{2}L$, $Z_{2}=2L+h$, $Z_{3}=h$, $Z_{4}=L+h$.
Numerical results for MFPs as a function of normalized wavevector are given in Fig.1 for a rectilinear arrangement of GBs together with the cases of SW defects and point impurities. The main model parameters are taken to be the same for all types of defects. 
\begin{figure} [tbh]
	\begin{center}
		\includegraphics [width=10.5 cm]{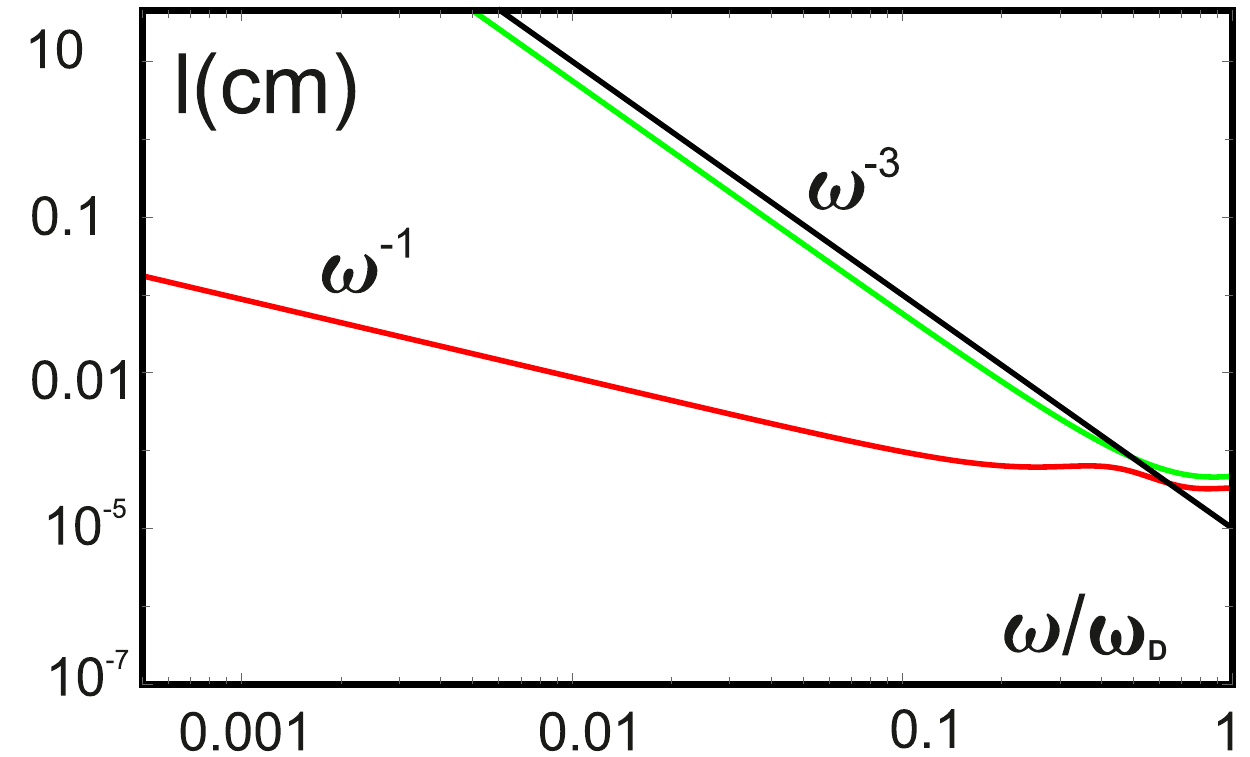}
	\end{center}
	\caption{The mean free paths of longitudinal phonons $l_{GB,L}$ (red line) and $l_{SW,L}$ (green line) as a function of $\omega /\omega _{D}$ with $\omega _{D}$ being the Debye frequency. The parameter set used is: $L=h=0.24$ nm, $\gamma _{LA}=1.7$, $n _{GB}=n _{SW}=2\times 10^{12}$ cm$^{-2}$, $\nu=0.16$, $\sigma =0.16$, $\delta=0.04$ nm.
	 For comparison, the MFP due to point impurity  is given (black line) with  concentration  $n _{pd}=2\times 10^{12}$ cm$^{-2}$ and the model parameters taken from Ref.~\cite{kras1}.}
\end{figure}

Two different regimes of  scattering are clearly seen in the long-wavelength limit: the MFP varies as $\omega ^{-3}$ for SW defects and point impurities while  $l_{GB}$  is found to be proportional to $\omega ^{-1}$ thus showing a dislocation-like behavior. What is important to note, we carried out calculations for GBs with different shapes (see two typical examples in Fig. 2) and can state that these two regimes are universal. 
\begin{figure} [tbh]
	\begin{center}
		\includegraphics [width=12 cm]{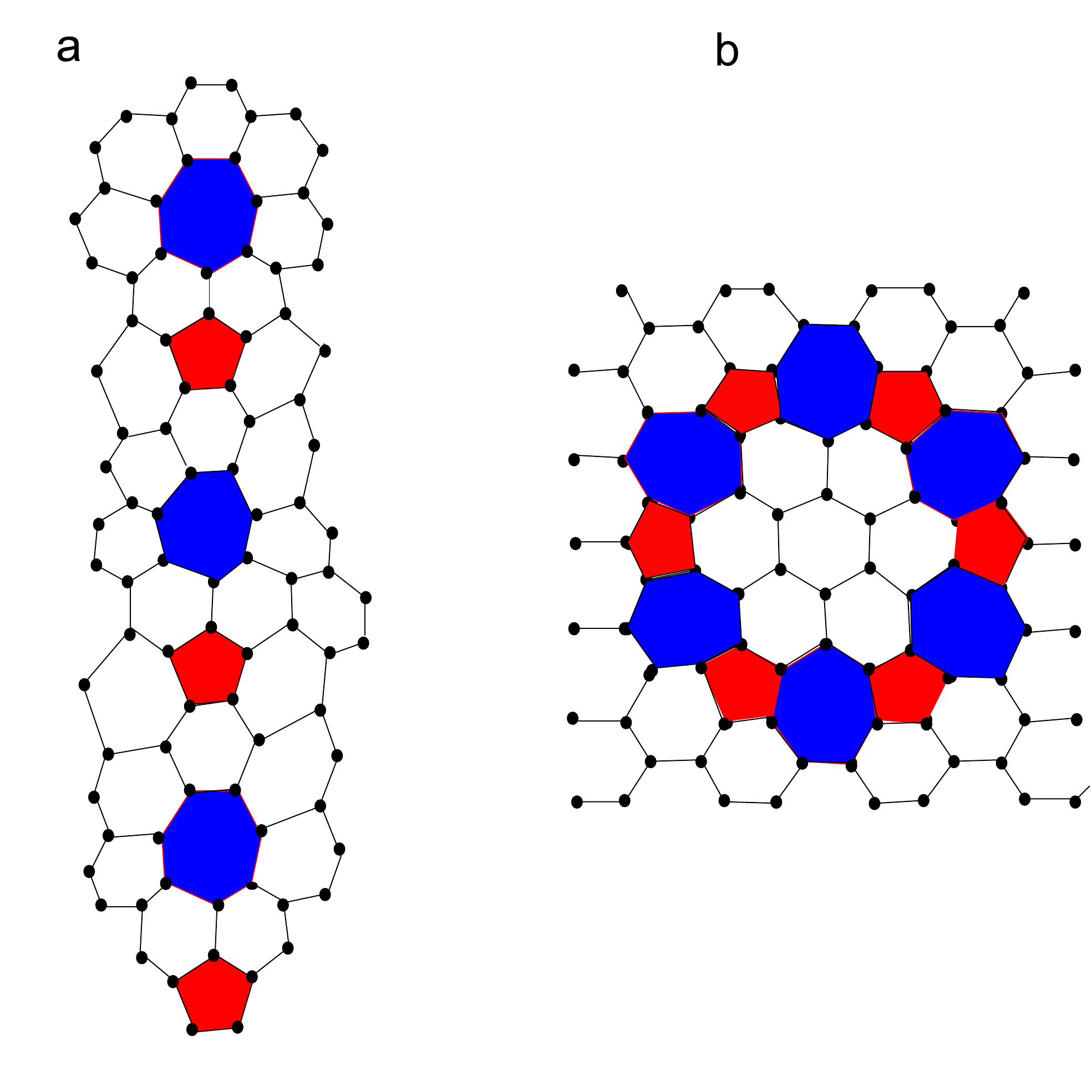}	
	\end{center}
	\caption{Two illustrative examples of GBs with different configurations in graphene: (a) open GB~\cite{wu1} and (b) closed GB~\cite{cock}.}
\end{figure}
Namely, any closed configurations show $\omega ^{-3}$ behavior like a point impurity. The diameter of the loop gives a characteristic size which defines the crossover point at large $\omega $ when the MFP starts to go to a constant.  
Notice that SW defect can be considered as a loop with a minimum radius. On the contrary, unclosed GBs always show $\omega ^{-1}$ behavior regardless of their shape. It has a simple physical explanation. Indeed, the strain fields caused by loops are strongly screened while they decay as $1/r$ for any unclosed configuration. The characteristic size of the GB (a shortest distance between end points) affects only the magnitude of the MFP and position of the crossover point. A shape of an open configuration is found to play a minor role.

There is an interesting specificity of GBs in graphene. To our knowledge, this is the first example when GBs show an internal structure. Let us recall that typically GBs are modeled by tightly packed walls of dislocations. In case of graphene the 5-7 disclination dipoles can be located at different (sometimes large enough) distances and, moreover, this determines the misorientation angle~\cite{rom,zhang}. At such an arrangement, the internal structure should manifest itself in the scattering of short-wave phonons. To clarify this, we carried out additional analysis of GBs  with a fixed length and different distances between dipoles (or, accordingly, different numbers $p$ of BWDDs  in the wall). The density of GBs is estimated as $n_{GB}=1/{\cal L}^2$ with ${\cal L}=pL+(p-1)h$ being the grain size and other model parameters are taken to be equal. The results are shown in Fig.3.
\begin{figure} [tbh]
	\begin{center}
		\includegraphics [width=10.5 cm]{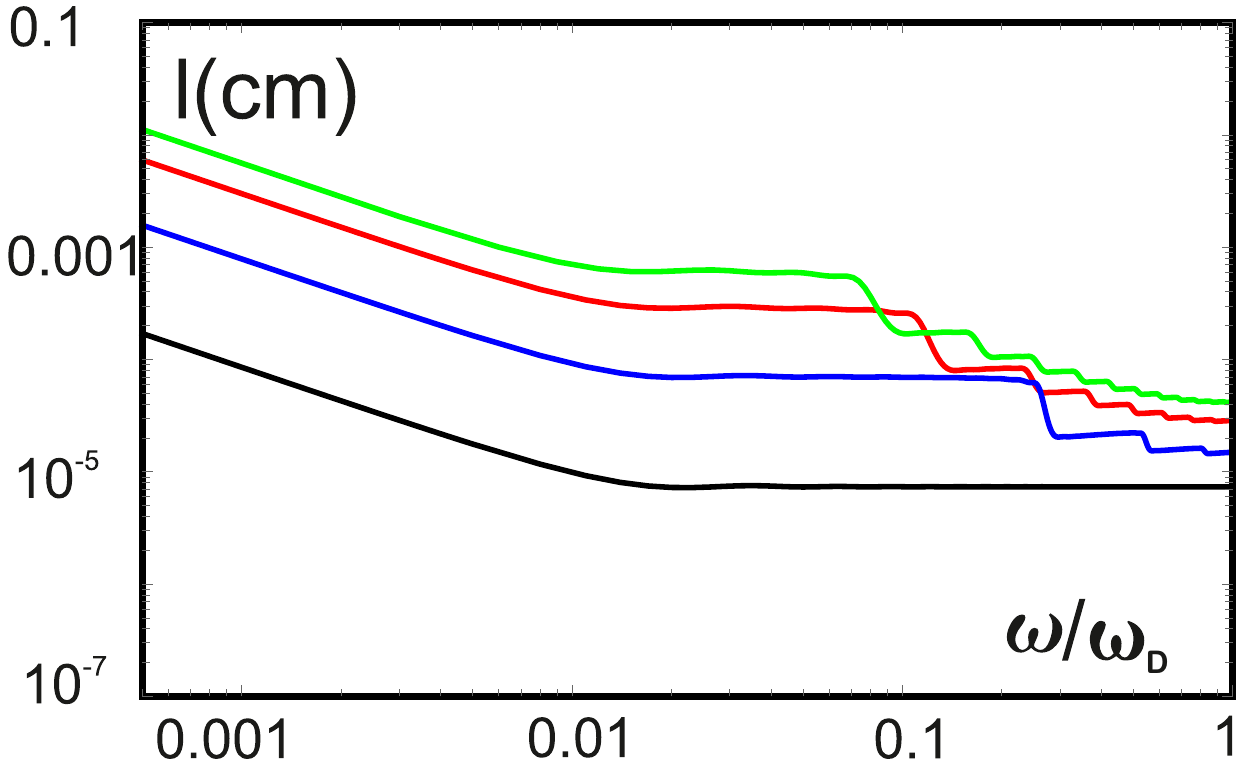}
	\end{center}
	\caption{The mean free paths of longitudinal phonons $l_{GB,L}$ as a function of $\omega /\omega _{D}$ with fixed grain size ${\cal L}=10$ nm and different distances between 5-7 pairs: $h=3.4$ nm (green line), $h=2$ nm (red line), $h=0.9$ nm (blue line). Black line corresponds to a single dipole of $10$ nm length. Other parameters are taken to be the same as in Fig.1.}
\end{figure}
As is seen, when phonon wavelength becomes comparable to $\cal L$, the crossover from $\omega ^{-1}$ to a constant behavior takes place. Then there appear steps in $l_{GB}(k)$ which are due to phonon diffraction on BWDD's wall. Actually, the GB acts as a one-dimensional periodic grating forming a reciprocal lattice in the  $k$-space with a period of $1/h$. The steps correspond to the Bragg's law. 
The value of $l_{GB}$ increases with increasing $h$ which is quite obvious: the wall becomes more transparent. Conversely, $l_{GB}$ takes the smallest value for a continuous wall with $p=1$ when the diffraction pattern completely disappears. We should note that for a periodic arrangement of dipoles the distance $h$ is directly related to the misorientation angle (see, e.g., Ref.~\cite{ovid1}): the greater $h$ the smaller the  misorientation angle. Therefore the GBs with larger misorientation angles will make a greater contribution to the thermal conductivity (see Fig. 3).

\section{Thermal conductivity}

We calculate the thermal conductivity of polycrystalline graphene by taking into account all main sources of phonon scattering. The total MFP is written as 
\begin{eqnarray}
1/l_{tot,\lambda }=1/l_{0}+1/l_{GB,\lambda }+1/l_{N,\lambda }+1/l_{U,\lambda },
\end{eqnarray}
where  $l_{0}$, $l_{GB,\lambda }$, $l_{N,\lambda } $ and $l_{U,\lambda }$ come from the phonon-rough boundary, phonon-GB,  three-phonon normal and Umklapp scattering processes, respectively, $l_0$ is an effective length determined from the geometry of graphene sample~\cite{alofi}, 
and 
 $l_{GB,\lambda}$ is given by Eq.(8). The MFP due to normal processes reads (see, e.g., Ref.~\cite{alofi,mor})
\begin{eqnarray}
l_{N,\lambda }^{-1}=B_{N,\lambda }\omega _{\lambda}^{2}(k)T^3,
\end{eqnarray}
and for Umklapp  phonon scattering we employ a parametrized expression in the form
\begin{eqnarray}
l_{U,\lambda }^{-1}=B_{U,\lambda }\omega _{\lambda}^{2}(k)T^3\exp(-\Theta _{\lambda}/3T).
\end{eqnarray}
Here $B_{N,\lambda }$ and $B_{U,\lambda }$ are model parameters, and $\Theta _{\lambda} $ is the Debye temperature. 

Within the Callaway's formalism, the diagonal components of the thermal conductivity tensor can be written as
\begin{eqnarray}
\kappa (T)=\kappa _{N}(T)+\kappa _{D}(T).
\end{eqnarray}
Here $\kappa _{N}(T)$ is the normal-drift term
\begin{eqnarray}
\kappa _{N}(T)=\frac{{\hbar ^2}}{2Sk_{B}T^2}\sum _{\lambda }\frac{\Bigl[\int d\omega \omega ^{2}_{\lambda}(k)l_{tot,\lambda}(\omega )l_{N,\lambda }^{-1}(\omega)v^{2}_{\lambda }C_{ph,\lambda }(\omega )N_{\lambda }(\omega )\Bigr]^{2}}{\int d\omega \omega ^{2}_{\lambda}(k)\Bigl(1-l_{tot,\lambda}(\omega )l_{N,\lambda }^{-1}(\omega )\Bigr)l_{N,\lambda }^{-1}(\omega )v^{3}_{\lambda }C_{ph,\lambda }(\omega )N_{\lambda }(\omega )},
\end{eqnarray}
where $l_{tot,\lambda}(\omega )$ is given by Eq.(9), $C_{ph,\lambda }(\omega )=\exp(\hbar \omega _{\lambda}(k)/k_{B}T)/(\exp(\hbar \omega _{\lambda}(k)/k_{B}T)-1)^2$, and $k_{B}$ is the Boltzmann constant. Summation is performed over all phonon polarization branches including out-of-plane (flexural) acoustic phonon branch (ZA). The explicit form of GB-induced MFP for flexural phonons ($l_{ZA}$) is taken from~\cite{kolesnik}, $N _{\lambda}(\omega )=S\omega _{\lambda}(k)/2\pi v_{\lambda}^{2}$ for $\lambda =LA,TA$, and $N _{ZA}(\omega )=S/4\pi b$ with $b$ being the bending elastic parameter (see Ref.~\cite{alofi}). The Debye thermal conductivity has the form 
\begin{eqnarray}
\kappa _{D}(T)=\frac{{\hbar ^2}}{2Sk_{B}T^2}\sum _{\lambda }\int d\omega \omega ^{2}_{\lambda}(k)l_{tot,\lambda}(\omega )v_{\lambda }C_{ph,\lambda }(\omega )N_{\lambda }(\omega ).
\end{eqnarray}
 Fig.4 shows  the thermal conductivity $\kappa (T)$ in a $2.9$ $\mu$m wide ribbon  containing  GBs with average sizes of $10$ nm and $2.5$ nm. The calculations have been done for 5-7-5-7 straight disclination wall. In order to estimate a contribution to the thermal conductivity from flexural phonons we carried out calculations with the maximum possible value of $l_{ZA}^{-1}$ (at $p=1$). We found that even in this case the contribution is insignificant with used parameter set. This agrees with the result of Ref.~\cite{kolesnik} where it was shown that ZA phonons manifest themselves only at very low temperatures.  
  \begin{figure} [tbh]
 	\begin{center}
 		\includegraphics [width=12.5 cm]{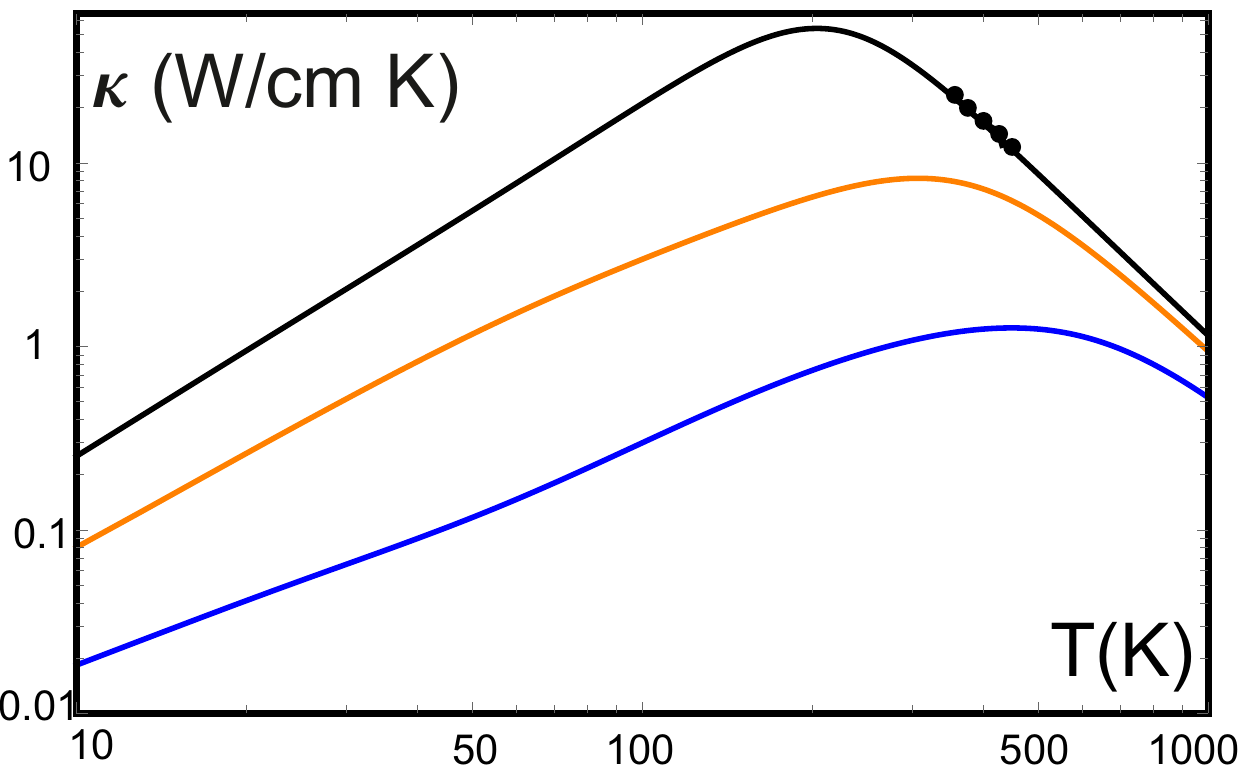}
 	\end{center}
 	\caption{ Thermal conductivity vs temperature for polycrystalline graphene with an average grain size of 2.5 nm (blue line, $h=0.3$ nm, $n_{GB}=1.6\times 10^{13}$ cm$^{-2}$ ) and 10 nm (orange line, $h=1.0$ nm. $n_{GB}=9.5\times 10^{11}$ cm$^{-2}$ ). GBs are considered as a row of 5-7 disclination dipoles. Other used parameters are: $l_0=2.9\times 10^{-4}$ cm, $B_{N,LA}=9.8\times 10^{-32}$s K$^{-3}$ cm$^{-1}$,  $B_{N,TA}=1.5\times 10^{-31}$s K$^{-3}$ cm$^{-1}$,  $B_{U,LA}=1.47\times 10^{-31}$s K$^{-3}$ cm$^{-1}$, $B_{U,TA}=2.27\times 10^{-31}$s K$^{-3}$ cm$^{-1}$,
 	$B_{N,ZA}=4.6\times 10^{-28}$s K$^{-3}$ cm$^{-1}$,
 	$B_{U,ZA}=6.9\times 10^{-28}$s K$^{-3}$ cm$^{-1}$.
 	 Black line corresponds to the case of pristine graphene.
Experimental points (circles) are taken from Ref.~\cite{chen}.}
 \end{figure}
 
One can see that $\kappa (T) $  markedly decreases with decreasing size of GBs. In our case, the smallest value of $\kappa (T)$  is found for GBs with a size of $2.5$ nm ($n_{GB}\approx 10^{13}$ cm$^{-2}$). It is interesting to note that the relation $\kappa (10$\mbox{nm}$)/\kappa (2.5$\mbox{nm}$)$ increases with temperature up to the maximum value of $\kappa (T) $. After that $\kappa (10$\mbox{nm}$)/\kappa (2.5$\mbox{nm}$)$ gradually goes to unity due to the increasing role of the Umklapp processes.
 Notice, that MD calculations provide a similar dependence of  $\kappa (T)/\kappa _{0}(T)$ for the same   grain sizes in the temperature range from $300 K$ to $1100 K$~\cite{wu}. 




 
 \section{Impact of irregularities and cracks} 

 As the experiment shows, in real samples GBs are serpentinelike and have some irregularities in the spatial arrangement of their structural blocks (heptagon-pentagon pairs)~\cite{huang,kim}. Thus the phonon-GB scattering in graphene becomes more complicated due to additional sources. According to the model suggested  in~\cite{ovid1,ovid2},  defects associated with elementary changes in the GB misorientation and their effects on crack generation are partial disclinations or disclination dipoles with strengths in the range of $-60^{\circ}<\tilde {\omega} < 60^{\circ}$ (see Fig.5). They appear inside GBs because of the step-like variations in the misorientations along GB lines. Generally, a partial disclination is defined as a point-like defect which can exist only at another line defect (a line GB) violating hexagonal topology outside the point-like core of this partial disclination (see Ref.~\cite{ovid2} for detail). Partial GB disclinations create stresses capable of initiating  nanocracks in graphene. In other words, these disclinations act as precursors of cracks. Thus, along with periodic GBs, we have additional sources of scattering due to either partial disclination dipoles (PDD) having coordinates $(x_{i2}-\eps,0)$ and $(x_{j1}+\eps,0)$  or they together with nanocracks  in accordance with  Fig.5.
  \begin{figure} [tbh]
  	\begin{center}
  		\includegraphics [width=14.0 cm,angle=360]{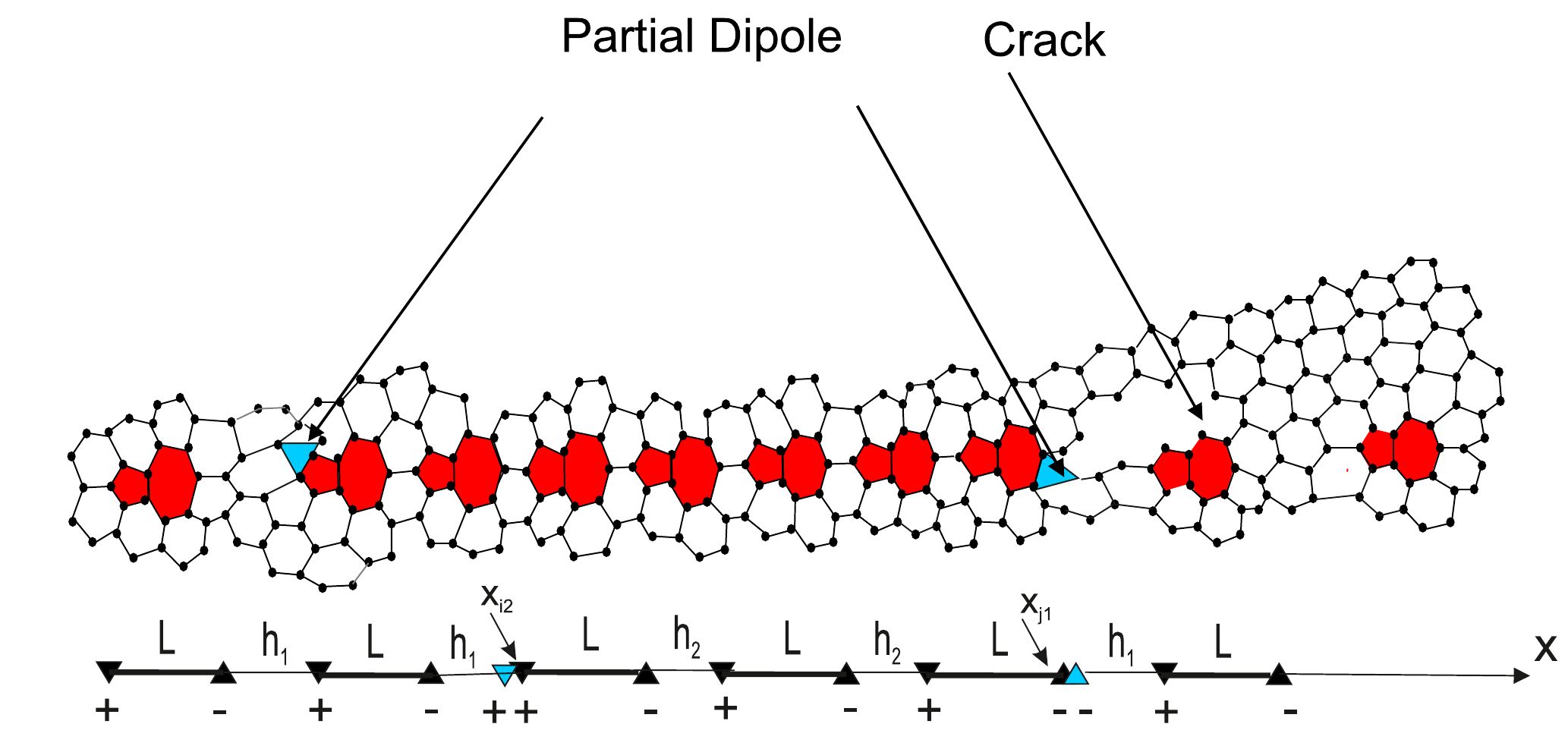}
  	\end{center}
  	\caption{Grain boundary having both changes in the misorientation angle and a nanocrack. The partial disclination dipole with the arm $2$ nm illustrated in figure by gray colour. It separated by seven heptagon-pentagon regular pairs shown by red colour. The power of partial dipole in calculations equal to $\pm 45^{\circ}$. Notice, that partial disclinations can be located in various places of GBs. Schematic picture below demonstrates a possible choice of dipole location.}
  \end{figure} 
 In the presence of PDD, the MFP can be written as
 $$
{\tilde l^{-1}_{GB,\lambda}}= l^{-1}_{GB,\lambda}+\frac{n_{PDD}\nu_{1}\nu _{2} D^2\pi }{2 k}\sum_{m=1}^{p}\Biggl(-\mathfrak{S}\biggl(k,x_{i2}-\eps,x_{m1},y_{i2},y_{m1}\biggr)+\mathfrak{S}\biggl(k,x_{j1}+\eps,x_{m1},y_{j1},y_{m2}\biggr)
 $$
 \begin{eqnarray}
 +\mathfrak{S}\biggl(k,x_{i2}-\eps,x_{m2},y_{i2},y_{m1}\biggr)-\mathfrak{S}\biggl(k,x_{j1}+\eps,x_{m2},y_{j2},y_{m2}\biggr) \Biggr)-\frac{n_{PDD}\nu ^{2}_{2}D^2\pi }{k}\mathfrak{S}\biggl(k,x_{j1}+\eps,x_{i2}-\eps,y_{j1},y_{i2}\biggr),
 \end{eqnarray}
 where $l^{-1}_{GB,\lambda}$ is the GB-induced  MFP of periodically ordered 5-7 pairs given by Eq.(6), $n_{PDD}$ is the density of PDDs, $\nu _{1}=\Omega/2\pi $,  $\nu _{2}=\tilde {\omega}/2\pi $.
 Figs. 6 and 7 show the MFP and normalized thermal conductivity for rectilinear GBs with $p=12$ and a length of  $6.5$ nm ($n_{GB}=2.3\times 10^{12}$ cm$^{-2}$) having built-in  dipole of partial disclinations  with the dipole arm of $2$ nm  and exactly one nanocrack per GB. 
   \begin{figure} [tbh]
  	\begin{center}
  		\includegraphics [width=12.5 cm]{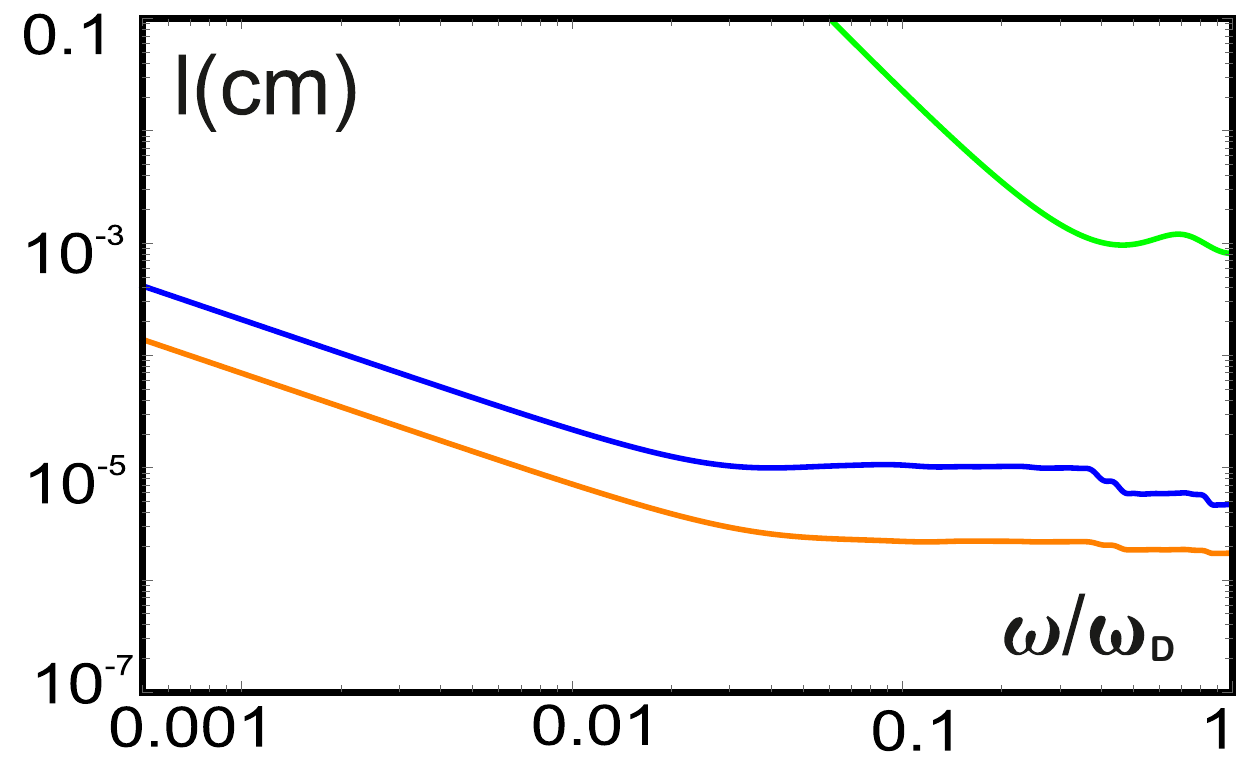}
  	\end{center}
  	\caption{The mean free paths of longitudinal phonons $l_{GB,L}$ as a function of $\omega $ for GBs of size ${\cal L}=6.5$ nm with an additional partial disclination dipole.  Distances between 5-7 pairs are chosen to be: $h=0.48$ nm (outside the PDD location) and $h=0.14$ nm (inside the PDD location). The PDD separation is taken to be $2$ nm.  The MFP for GB without PDD (blue), with PDD (orange), and for a nanocrack (green). The model parameters are: $n_{cr}=n_{GB}=n_{PDD}=2.3\times 10^{12}$ cm$^{-2}$, the strength of the PDD $\omega =45^{\circ}$,  $p=12$. Other  model parameters are taken to be the same as in Fig.3. }
  \end{figure}
  \begin{figure} [tbh]
   	\begin{center}
   		\includegraphics [width=12.5 cm]{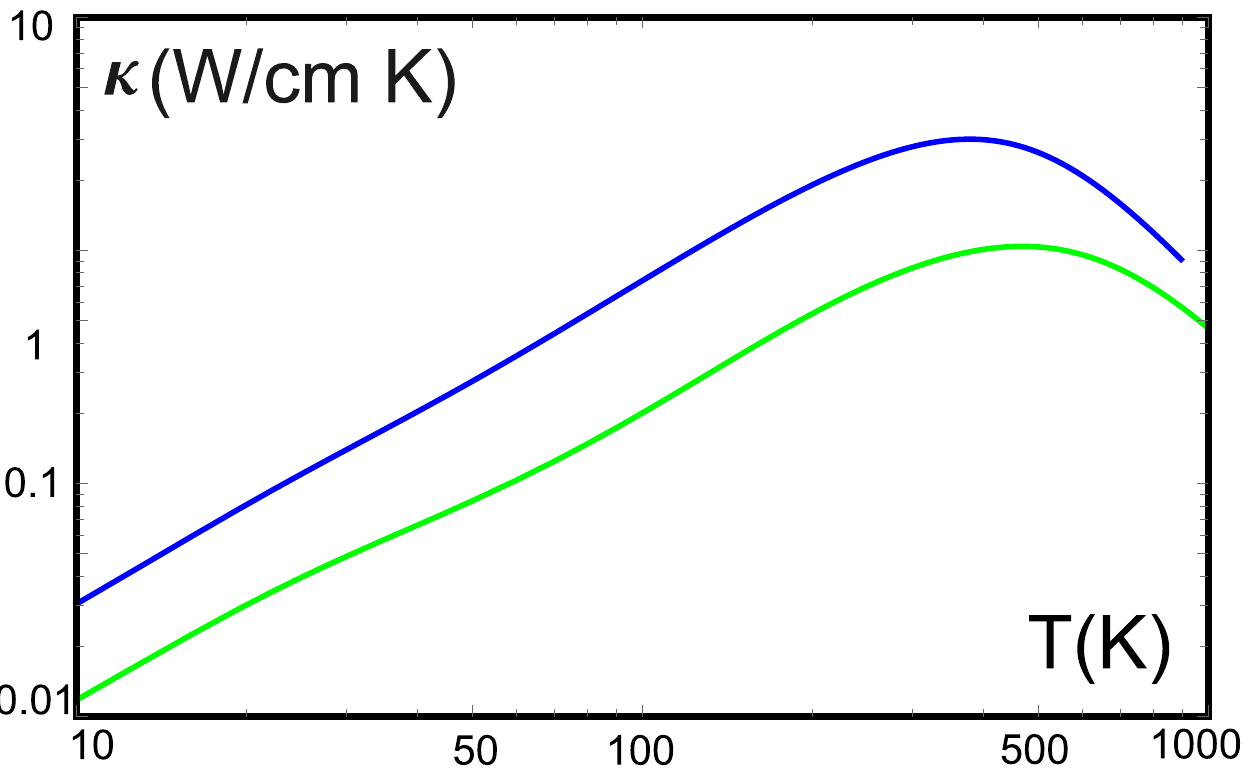}
   	\end{center}
   	\caption{ Thermal conductivity vs temperature for polycrystalline graphene with an average grain size of 6.5 nm and a partial dipole of wedge disclinations inside GB $2$ nm in length and the strength $\tilde {\omega} = 45^{\circ}$ (green). Blue line corresponds to GB without partial dipole. For both curves the density of defects is  $n_{GB}=2.3\times 10^{12}$ cm$^{-2}$. Other  parameters are the same as in Fig.4.}
   \end{figure}
This case corresponds to the maximum possible density of PDDs and nanocracks $n_{cr}$ when $n_{cr}=n_{GB}=n_{PDD}$. 

Fig. 6 shows the phonon MFP for GBs both with and without the PDD. As is seen, with the same model parameters, the MFP for GBs containing the PDD lies noticeably lower in comparison with that for a conventional GB. There is a difference of about 3 times. For clarity, this can be derived from the expansion of the MFP in Eq. (15)  when $k$ tends to zero. One obtains $l_{GB+PDD}/l_{GB}\sim (1+\tilde {\omega} d/\Omega pL)^2$ where $d$ is the arm of the PDD. The last term in the brackets is always less than unity, so that the maximum possible reduction in the MFP is four times. Notice also that oscillations in the region of large wave vectors exist but slightly depressed. Evidently, we are faced here with increased stresses and the destruction of periodicity due to PDD.

Separately, we considered the GB containing PDD combined with a nanocrack. In order to calculate a contribution from cracks one can use an analogy with $3D$ phonon scattering due to platelets (see, e.g., Ref.~\cite{turk}). In the two-dimensional case, the MFP is found to be proportional to $\omega ^{-3}$ in the long-wavelength limit like for point imperfections or SW defects. At large $\omega $, the phonon MFP tends to a constant  (see Fig. 6). On the whole,  MFP for nanocracks is similar to that for SW defect with the only difference in a larger characteristic size. Which is important, the MFP curve lies for nanocracks much higher than those for GBs  (blue and orange lines) for all $\omega $ even at critical (practically unreachable) density $n_{cr}=n_{GB}=2.3\times 10^{12}$ cm$^{-2}$ and at chosen nanocrack size of $1$ nm. It means that this scattering channel cannot noticeable affect $\kappa (T)$. 

 
Fig.7 shows the thermal conductivity of polycrystalline graphene $\kappa (T)$ without PDD and with PDD of length $L=2$ nm and strength $\tilde {\omega} = 45^{\circ}$. It is seen that the thermal conductivity for GB with PDD decreases noticeably in a wide temperature range up to high temperatures. This agrees with the behavior of MFP in Fig. 6. Additionally, it means that although the role of Umklapp processes increases with temperature, they still do not become dominant. The decrease in thermal conductivity increases with increasing arm and/or the power of the PDD. Our calculations show that an effect of nanocracks on the total thermal conductivity is really insignificant for the chosen model parameters.

\section{Conclusion} 

In this paper, we investigated theoretically the influence of GB characteristics on the heat transport in polycrystalline graphene within the Callaway approach where normal phonon processes have been included.
GBs have been considered as rows of repeated $5-7$ topological dislocations or disclination dipoles. What is important, GBs of any configuration can be described using the developed  formalism. We have analyzed the following examples: the straight GB of finite length, the Stone-Wales defect (two dipoles forming rhombus), and straight GBs with partial disclination dipole inside the wall both with and without nanocrack. It is found that the phonon MFP behaves in two ways in the long-wavelength limit: for open GBs it varies as $\omega ^{-1}$, while for closed configurations as $\omega ^{-3}$. For short waves, the MFP is $\omega $-independent for all configurations of GBs. We have shown that the existing internal structure of linear GBs in graphene leads to diffraction of phonons for short waves at large enough number of 5-7 dipoles in the wall. This clearly manifests itself on the MFP curves  in the form of step-like oscillations (see, for example, Fig.3). The presence of partial disclination dipoles, which appear in the GBs when a misorientation angle varies, drastically changes the picture. In this case, we have obtained a noticeable decrease of the thermal conductivity. Our calculations show that nanocracks inside GBs with PDDs have a minor impact and do not affect the total MFP even at their abnormally high density (one crack per GB). We found that the thermal conductivity is much more sensitive to the internal structure of GBs rather than to their geometrical shape and decreases in a wide temperature range with an increase in the angle of misorientation. A noticeable drop in thermal conductivity was obtained when additional partial disclinations associated with changes in the misorientation angles are present inside GBs. 

Finally, our study confirms the previously obtained theoretical results as well as recent experimental observations~\cite{vlassiouk, ma2, lee} showing a significant decrease in the thermal conductivity of polycrystalline graphene. Moreover, we have shown that this conclusion is valid in a wide temperature interval and is practically independent of GB's shape. For example, it holds for wriggling grain boundaries. Irregularities in real structures violating the ideal constancy of the GB misorientation lead to a further marked reduction of the thermal conductivity. This is important to take into account when designing thermoelectric devices based on polycrystalline graphene through grain boundary engineering.



	
\appendix

\section{Dilatation strain tensor of a grain boundary}

The stress field of arbitrary GB lying in the $xy$-plane can be obtained as a sum of stresses of $p$  wedge disclination dipoles of the same strength $\Omega $ and fixed separation $L$ (see Refs.~\cite{Li,hurt})
\begin{equation}
\sigma _{xx}^{(d)}(r)+\sigma _{yy}^{(d)}(r)=\frac{G\Omega }{2\pi(1-\sigma)}\sum_{n=1}^{p}\ln\frac{(x-x_{n1})^2+(y-y_{n1})^2}{(x-x_{n2})^2+(y-y_{n2})^2},
\end{equation}
\begin{equation}
\sigma_{zz}^{(d)}(r)=\frac{G\Omega \sigma }{2\pi(1-\sigma )}\sum_{n=1}^{p}\ln\frac{(x-x_{n1})^2+(y-y_{n1})^2}{(x-x_{n2})^2+(y-y_{n2})^2},
\end{equation}
where $G$ is the shear modulus, $\sigma $ is Poisson's ratio, $(x_{ni},y_{ni})$ are coordinates of $i$-th disclination in $n$-th dipole. If dipoles are oriented along the $x$ axis, the coordinates $x_{ni}$ should satisfy the condition  $|x_{n2}-x_{n1}|=L$ in Eqs. A(1) and A(2). The explicit expression for dilatation $TrE_{ij}^{d}(r)$ in Eq.(3) may be found from Eqs. A(1) and A(2) and Hook's law as
$$
TrE_{ij}^{d}(r)=\frac{(1-2\sigma )}{2G(1+\sigma)}Tr\sigma _{ij}^{d}(r)
$$
\begin{equation}
=\frac{(1-2\sigma ) }{4\pi(1-\sigma)}\left(\Omega\sum_{n=1}^{p}\ln\frac{(x-x_{n1})^2+(y-y_{n1})^2}{(x-x_{n2})^2+(y-y_{n2})^2}+
\omega \ln\frac{(x-x^{\prime}_{m2})^2+(y-y^{\prime}_{m2})^2}{(x-x^{\prime}_{m1})^2+(y-y^{\prime}_{m1})^2}\right).
\end{equation}
The last term in Eq.A(3) describes the dilatation for partial 
disclination dipole with a power of $\mp\omega $ located at points $(x^{\prime}_{m1(2)},y^{\prime}_{m1(2)})$.
The function $\mathfrak{S}\biggl(k,x_{ni},x_{mj},y_{ni},y_{mj}\biggr)$ in Eq.(6) has the following explicit form: 
$$
\mathfrak{S}\biggl(k,x_{ni},x_{mj},y_{ni},y_{mj}\biggr)=-2\pi k^{2}\Biggl(\biggl(x_{ni}-x_{mj}\biggr)^2+(y_{ni}-y_{mj}\biggr)^2 \Biggr)
$$
$$
\Biggl(J_{0}^{2}(k\sqrt{((x_{ni}-x_{mj})^2+(y_{ni}-y_{mj})^2)}+J_{1}^{2}(k\sqrt{((x_{ni}-x_{mj})^2+(y_{ni}-y_{mj})^2)}
$$
$$
-\frac{1}{k\sqrt{((x_{ni}-x_{mj})^2+(y_{ni}-y_{mj})^2)}}
$$
\begin{eqnarray}
J_{0}(k\sqrt{((x_{ni}-x_{mj})^2+(y_{ni}-y_{mj})^2)}J_{1}(k\sqrt{((x_{ni}-x_{mj})^2+(y_{ni}-y_{mj})^2)} \Biggr).
\end{eqnarray}

This is a result of integration of Eq.(1) with Eq.(5) where $q=|k-k'|=2k\sin (\theta/2)$ (see also details in Ref.~\cite{kras3}).

\end{document}